%% file: main.tex
\documentclass[sigconf]{acmart}

\AtBeginDocument{%
  \providecommand\BibTeX{{%
    \normalfont B\kern-0.5em{\scshape i\kern-0.25em b}\kern-0.8em\TeX}}}

\copyrightyear{2025}
\acmYear{2025}
\setcopyright{acmlicensed}\acmConference[CHI '25]{CHI Conference on Human Factors in Computing Systems}{April 26-May 1, 2025}{Yokohama, Japan}
\acmBooktitle{CHI Conference on Human Factors in Computing Systems (CHI '25), April 26-May 1, 2025, Yokohama, Japan}
\acmDOI{10.1145/3706598.3713988}
\acmISBN{979-8-4007-1394-1/25/04}

\begin{document}

\title[WhatsApp for Business]{"Business on WhatsApp is tough now—but am I really a businesswoman?" Exploring Challenges with Adapting to Changes in WhatsApp Business}


\author{Ankolika De}
\email{apd5873@psu.edu}
\affiliation{%
  \institution{College of Information Sciences and Technology, Pennsylvania State University}
  \country{USA}
}
\renewcommand{\shortauthors}{De}

\begin{abstract}

This study examines how WhatsApp has evolved from a personal communication tool to a professional platform, focusing on its use by small business owners in India. Initially embraced in smaller, rural communities for its ease of use and familiarity, WhatsApp played a crucial role in local economies. However, as Meta introduced WhatsApp Business with new, formalized features, users encountered challenges in adapting to the more complex and costly platform. Interviews with 14 small business owners revealed that while they adapted creatively, they felt marginalized by the advanced tools. This research contributes to HCI literature by exploring the transition from personal to professional use and introduces the concept of \textit{Coercive Professionalization}. It highlights how standardization by large tech companies affects marginalized users, exacerbating power imbalances and reinforcing digital colonialism, concluding with design implications for supporting community-based appropriations.

\end{abstract}



\maketitle

\input{introduction}

\input{background}

\input{methods}

\input{findings}

\input{discussions}

\input{limitations}

\input{conclusionandfuture}


\bibliographystyle{ACM-Reference-Format}
\bibliography{references.bib}



\end{document}

%% file: introduction.tex
\section{Introduction}

Businesses have increasingly leveraged messaging applications to enhance various aspects of their operations \cite{anderson2016getting, lo2022mobile, doi:10.1177/20501579241246721}. Specifically, messaging applications like WeChat \cite{yang2016role}, WhatsApp \cite{10.1145/3613905.3651034}, Facebook Messenger and Telegram \cite{LoPresti2021}  have proven invaluable for business communication, aiding in customer acquisition, retention, and overall marketing efforts. Within HCI (Human-Computer Interaction) literature, scholars have investigated the various ways in which communication technologies have been appropriated by communities to empower themselves (see for eg., \cite{10.1145/2660398.2660427, 10.1145/3328020.3353927, 10.1145/3318140, 10.1145/2145204.2145220}). 

Before Meta introduced WhatsApp Business as a professionalized platform, WhatsApp was already widely used for business purposes. In India, particularly in smaller, rural areas \cite{doi:10.1177/20501579241246721} and financially constrained communities \cite{10.1145/3014362.3014367} with limited technical resources, WhatsApp became a crucial tool for conducting business and supporting local economies \cite{modak2017dancing}. Additionally, the platform's broad use across different age groups, including older adults, also spoke for its important role and impact \cite{10.1145/3449212}. However, as its appropriation for business became widespread, Meta introduced a formalized version of WhatsApp Business. This update included changes to the interface and additional costs. Alongside these changes, Meta began regularly altering WhatsApp’s features and functionality \cite{10.1145/3613905.3651034}. For example, features like account verification, message-based pricing, and business verification for accessing full features became necessary to appear as legitimate \cite{meta_verified_business_2024}.

While change is inevitable and often driven by policies, design updates, and business strategies, it remains understudied within the HCI community to understand its impact on communities who have successfully appropriated technologies for empowerment \cite{10.1145/3443686, 10.1145/3613905.3651034}. It is important to investigate how platforms, initially shaped and \textit{infrastructured} through user contributions \cite{10.1145/3313831.3376201}, are later co-opted by larger corporations \cite{doi:10.1177/1461444816629474}. These corporations often modify the platforms, shifting control and benefits away from the original users, which exemplifies digital colonialism \cite{Thorat2020}. Thus, I am interested in understanding how the consequent rapid changes in technology impact traditionally decentered users' ability to understand and adapt. This focus is crucial for designing and implementing changes inclusively, which are key topics in HCI \cite{10.1145/3544548.3581040}. Thus, in this study, I examine how WhatsApp users—whether they use personal accounts or WhatsApp Business accounts—adapt to changes on the platform. I start by exploring how these users initially adopted WhatsApp for business purposes and then analyze their responses and experiences with subsequent changes to the platform. Thus, I ask:

\textbf{RQ1: How did users initially transition from using WhatsApp as a personal communication tool to adopting it for business purposes?}
    
\textbf{RQ2: What impact did the frequent changes and updates in the WhatsApp Business platform have on users' business operations and how did they adapt to these changes?}

I conducted interviews with 14 users who employed WhatsApp for home-based small businesses in India and subsequently applied thematic analysis to address the research questions. The findings showed that users leveraged the \textit{personal nature of WhatsApp} to rely on their networks and personal experiences to adapt WhatsApp for business purposes. Despite experiencing confusion and challenges in adapting to changes, participants often devised innovative solutions to maintain their familiarity with the platform. However, they also often trivialized the changes, feeling \textit{undeserving} of such tools due to their perceived marginalized status and their belief that they were not the platform’s primary target users. They perceived these updates as more beneficial for larger businesses, pointing to various new features released by Meta as part of WhatsApp Business that were aimed at enhancing business professionalism.

I contribute to HCI literature in several ways. First, I discuss how personal communication tools become important spaces where innovation and community-based appropriations occur through collaborative processes. These informal settings provide a comfortable environment for many users to experiment and adapt tools in creative ways, as exemplified by WhatsApp Business. Consequently, I explain WhatsApp Business as an infrastructuralized platform \cite{doi:10.1177/1461444816661553} and use the appropriation matrix \cite{10.1145/3613904.3642590} to document how its business features, initially a local innovation, have been \textit{reverse-appropriated} by Meta \cite{10.1145/3613904.3642590}. I then introduce \textit{Coercive Professionalization}, which explains how large tech companies, such as Meta with WhatsApp, monetize and formalize informal practices originally developed by local users. Drawing on Lampland and Star's \cite{lampland2009standards} work on \textit{standards}, I explore how they impose normative categories on messy real-world situations, and how participants present their feelings and understandings of changes they see as attempts to appease larger businesses, often with a sense of \textit{resigned acceptance}. Finally, drawing from De's \cite{10.1145/3613905.3651034} work on \textit{situated infrastructuring}, I analyze how, combined with coercive professionalization, it reinforces colonial practices, exacerbating power imbalances and marginalization in ways that may not be apparent to those affected.

%% file: background.tex
\section{Background}

\subsection{WhatsApp in India: A Brief Review}

\begin{table*}[t]
\footnotesize
  \caption{ \label{tab: Changes} Examples of WhatsApp Business' platform changes since its deployment \cite{whatsapp_business_pricing_updates_2024}; these changes are ongoing and being rolled out in different locations and for individual users at various times. }
  \label{tab:bus}
\resizebox{\textwidth}{!}{%
\begin{tabular}{p{6cm} p{6cm}} 
   \toprule
   \textbf{Original Plan} & \textbf{Changed State} \\
   \midrule
   
   \textbf{Pricing Model (Per Conversation)}: Businesses were charged per conversation, regardless of the number of messages. & \textbf{Pricing Model (Per Message)}: Businesses are now charged per message, with different rates for marketing, utility, and service messages. \\   \textbf{Utility Template Charges}: Utility templates outside the customer service window—defined as the designated hours during which customer support is available—were previously charged per conversation.& \textbf{Utility Template Charges}: Utility templates are now charged per message, rather than per conversation, when sent outside the service window.\\   
   \textbf{Entry Point Conversations (Free Service Window)}: Entry point conversations allowed for a free service window, but were counted as a conversation. & \textbf{Entry Point Conversations (Free Service Window)}: Entry point conversations are free for 24 hours, and then open a customer service window lasting 72 hours. \\
   
  \textbf{Authentication Rate Eligibility (verifying the identity of a business account)}: Eligibility for international authentication rates was previously based on the number of conversations opened. &\textbf{Authentication Rate Eligibility}: Eligibility is now based on the number of authentication messages sent, rather than the number of conversations opened.\\ 

   \textbf{Business Verification (Earlier Context)}: Meta business verification was required, but businesses could proceed with limited features without it. & \textbf{Business Verification (Current State)}: Business verification is a norm now, with several subscription plans that businesses must choose from to appear reliable and use all functionalities of the application. \\
   
   \bottomrule
\end{tabular}%
}
\end{table*}

Founded in 2009 and now owned by Meta, WhatsApp is the world's most used messaging app \cite{dixon2024}, with India being its largest user base \cite{nair2024}. What makes WhatsApp so popular is both its lack of complexity, as well as the relatively low barrier of participation that it expects \cite{10.1145/3014362.3014367, Pang2020, Garimella2024}. This helps \textit{emergent users} overcome complexity barriers, allowing them to engage with the platform more effectively than with more complicated alternatives \cite{10.1145/3014362.3014367, 10.1145/2493190.2493225}. WhatsApp appeals to the masses by offering flexible, user-preference-based communication without algorithmic mediation, providing a convenient and personal way to connect with distant contacts \cite{10.1145/3512964, 10.1145/2531602.2531679}. 

Researchers have highlighted WhatsApp's ritualized role in India's communication landscape, referring to Couldry's \cite{rothenbuhler2005media} concept of ritualization, which encourages researchers to examine the connections between ritual actions and broader social contexts, including the practices, beliefs, and categories that make these rituals possible. This helps explain why people prefer it over other tools for maintaining their social lives \cite{doi:10.1177/01968599221095177}. The following sections highlight WhatsApp's unique content, its evolution and adaptation as a business tool, and how localized practices in the Global South enhance the effectiveness of WhatsApp Business. I emphasize the need to study how marginalized users adapt WhatsApp Business in culturally specific ways. This research aims to address gaps in the literature by examining how marginalized groups adapt to changes, proposing improvements to better support them, and contributing to strategies that prevent their decentering in design processes. Here, decentered users refer to individuals or communities who, due to broader systemic forces and technologies, are systematically displaced from the center of design considerations, often leading to their needs being overlooked or inadequately addressed \cite{belfer_center_design_from_the_margins}. This research emphasizes the importance of repositioning these users at the core of design processes to ensure more inclusive, equitable, and effective outcomes \cite{belfer_center_design_from_the_margins_2}.

\subsubsection{WhatsApp's Unique Content Sharing Practices}
Unlike Twitter and Facebook, where original posts remain intact when shared, WhatsApp often fragments and disperses content without retaining the original sources \cite{10.1145/3512964}. This informality and ease of information sharing also make it highly susceptible to misinformation and disinformation \cite{paris2024hidden, 10.1145/3641010, 10.1145/3432948}. Additionally, it poses challenges related to spam, information overload, and other concerns prevalent in digital applications \cite{shahid2024one}. While tools to combat misinformation are widely studied, scholars also note that community-based code-switching helps assess trustworthiness and information credibility \cite{10.1145/3637429}. In India, the ritualized use of WhatsApp frequently hinges on such community-based, trust-oriented interactions \cite{10.1145/3613905.3651034, 10.1145/3491102.3517575}, where personal connections play a central role even in professional settings \cite{Durgungoz2022, 10.1145/3411764.3445221}. Business on WhatsApp, where personal and professional uses are intertwined, represents a unique techno-cultural phenomenon that warrants further investigation.

\subsubsection{WhatsApp for Business: An Evolution}

Using WhatsApp for business represents a culturally unique yet natural adaptation of the platform, showcasing how it empowers users in resource-constrained settings and warrants further investigation \cite{10.1145/3555584}. Users have adopted fragmented and informal methods for business operations, including creating groups and communities to engage customers, facilitate transactions, and leverage word-of-mouth for promotion \cite{10.1145/3613905.3651034, Sugiyantoro2022, modak2017dancing, kottani2021study, bagdare2021whatsapp}. Particularly, during the pandemic, online business use surged as in-person stores, often trusted by customers, had to close temporarily or shut down \cite{sumarni2022utilization, Sugiyantoro2022}. Many businesses transitioned to digital modes maintain connections with their loyal customers and continue operations virtually \cite{doi:10.1177/20501579241246721}. However, a critical gap remains in understanding how the appropriation of a personal messaging tool into a business tool occurs. This study aims to address this gap using WhatsApp as a case example.

WhatsApp for Business differs from the personal version by offering features tailored for business use. While WhatsApp (personal) is designed for private communication, WhatsApp for Business includes tools like customizable business profiles, product catalogs, automated messaging, and quick replies to streamline customer interactions. Additionally, businesses can run ads that direct users to WhatsApp chats, a feature not available in the personal app. Overall, WhatsApp for Business is designed to support professional communication and scalability, unlike the personal app \cite{WhatsAppPricing2024}.

WhatsApp Business, originally designed to help businesses of all sizes communicate with customers through features like business profiles, catalogs, and automated responses, has gradually evolved into a more professionalized platform \cite{whatsappforbusiness} (Also see table \ref{tab: Changes} that details selected changes). Meta's shift to per-message pricing for WhatsApp Business, following earlier per-conversation pricing changes, significantly increased costs for small businesses, prompting some to explore alternatives \cite{10.1145/3613905.3651034}. This shift meant that businesses were charged for each individual message sent, including marketing, authentication, and utility templates outside of the customer service window, which increased costs for those with large-scale messaging operations \cite{cornish2023whatsapp}. Other recent updates, including pricing changes, verification process, and new messaging features, also signal a shift toward a more enterprise-level platform \cite{WhatsAppPricing2024}. For small businesses, this leads to higher costs, especially for customer acquisition. It also introduces other complexities like enhanced payment options which can be activated by linking a bank account and using UPI (Unified Payments Interface), a real-time payment system that enables instant fund transfers, with payments completed after reviewing the order and entering a UPI code for authentication \cite{whatsapp_payment_guide_2024}. However, these changes have made the platform more expensive and are perceived as less accessible for smaller businesses \cite{10.1145/3613905.3651034}.

Drawing from prior research on the rhetoric of WhatsApp Business and its formal introduction in 2018—following users' prior appropriation of personal WhatsApp accounts for local, community-centric purposes—this study examines users' adaptive strategies and responses to platform updates \cite{10.1145/3613905.3651034}. It empirically explores how marginalized business owners, who may face challenges due to limited digital skills, use WhatsApp Business to maintain and grow their operations. Studying this phenomenon is crucial for the HCI community because it sheds light on how digital tools impact marginalized groups and highlights the need for design practices that accommodate diverse user contexts-- so as to not decenter them through design.  By understanding the adaptive strategies of these users, this research can contribute to the development of more inclusive and supportive technologies that enhance successful appropriation and empowerment for all users, particularly those with limited digital resources.

Thus, I aim to bridge the gap in existing literature by examining how marginalized users—particularly those from resource-constrained settings—adapt WhatsApp Business to meet their specific needs. These users, often overlooked in the design and development of digital tools, face unique challenges that hinder their ability to fully utilize evolving technologies. By exploring how these users navigate frequent platform updates and integrate WhatsApp Business into their operations, the study seeks to offer insights into more inclusive design practices. Centering them can contribute to the development of more accessible, equitable, and user-centered digital tools that empower marginalized communities and ensure their needs are addressed in future technology design.

\subsection{Appropriating Infrastructuralized Platforms}

Platforms are digital systems that enable users to interact, create content, or exchange services, often with some level of reprogrammability or customization (e.g., social media platforms, app stores) \cite{doi:10.1177/2056305115603080, 10.1145/3313831.3376201, doi:10.1177/1461444816661553}. In contrast, infrastructures are large-scale foundational systems that support other services or systems, typically serving a broad range of users and becoming embedded in everyday practices (e.g., the internet, electricity grids) \cite{star1996steps}. As has been discussed in prior work, Meta’s global network of interconnected platforms, through its control over communication channels and data infrastructure, functions as a critical communication infrastructure, supporting various business operations, user interactions, and content delivery worldwide \cite{lunden2024meta}. Early views of infrastructure were constrained to physical, technical, and material systems, such as railways and power grids, which were seen as foundational \cite{article2006}. Contemporary research \cite{star1996steps} views infrastructure as a sociotechnical and relational phenomenon rather than merely a supportive framework, which is the perspective I adopt in this work. My focus is on internet infrastructures, which also develop gradually and are influenced by a range of social, political, and economic factors \cite{doi:10.1177/14614448231152546}. 

Platforms evolve into infrastructures when they extend beyond their initial purpose, become deeply embedded in everyday life, and function as foundational elements for other systems. This transformation often occurs through user appropriation and ongoing processes of infrastructuring-- defined as \textit{"the intentional production of infrastructure as a means of achieving a particular goal or the desire to solve a particular problem"} \cite{10.1145/3359175, 10.1145/2818048.2820015, doi:10.1177/0162243913516012}. Thus, platforms evolve from providing specific services to becoming essential, ubiquitous components of the broader technological and social landscape \cite{doi:10.1177/1461444816661553}-- and users' appropriation is often a contributor to the creation of infrastructuralized platforms-- as explicated in \cite{10.1145/3313831.3376201} about WeChat. I draw from Stevens \cite{stevens2009appropriation}, who conceptualizes appropriation as an "\textit{ongoing process}" in which users adapt and transform a technology (in my case a platform) to suit their context, simultaneously shaping its functionality and meaning. This shift allow platforms to become default choices, enabling their owners—such as corporations or service providers—to maintain their original reputations and collect more data from users \cite{10.1145/3313831.3376201}. 

In this study, I use infrastructuring as a dynamic lens where users continuously adapt \cite{karasti2018studying} and refine their digital business practices using WhatsApp, tailoring the platform to meet their specific needs and changing circumstances. Once platforms become infrastructure, their owners gain substantial power over users due to limited alternatives, causing users to continually adjust their \textit{folk theories} to adapt to ongoing changes \cite{10.1145/3476080, 10.1145/3392847, 10.1145/3533700, 10.1145/3359146}. When platforms like WhatsApp are vital to livelihoods, such as for small businesses, users engage in what can be described as \textit{routine infrastructuring} \cite{10.1145/3359175}. This involves being resilient and adapting to frequent disruptions within the platform, which requires ongoing effort to maintain and adjust the infrastructure.


Mwesigwa and Csíkszentmihályi \cite{10.1145/3613904.3642590} developed the appropriation matrix to explain \textit{appropriation} (initial introduction of technologies by companies and governments), \textit{re-appropriation} (secondary use and appropriation of technologies by local communities and development actors), and \textit{reverse appropriation} (formal integration of these uses into more professional and often unaligned systems by companies). This can be explained through WhatsApp's lifecycle. Initially, WhatsApp was introduced as a communication tool for personal messaging (appropriation). However, local businesses, especially small enterprises, began using WhatsApp not just for messaging, but also for customer service, order processing, and even marketing (re-appropriation)—which is the focus of my first research question. This use was not part of the original design, but it became so widespread that WhatsApp eventually introduced business-specific features, such as WhatsApp Business accounts, which allow businesses to manage customer interactions more professionally (reverse appropriation). The adaptation to these changes is the focus of my second research question. 

Users of infrastructural platforms for purposes such as care \cite{10.1145/3544548.3581040}, networking \cite{Crabu2018}, protection \cite{10.1145/3313831.3376339}, and well-being \cite{wilson2016infrastructure} may encounter alterations in these systems that create vulnerabilities, often unnoticed and overlooked for profit-driven reasons due to reverse appropriation \cite{Hague2019, 10.1145/3476080}. This is evident in how Meta, for example, capitalizes on users' appropriated systems, such as their use of WhatsApp for business, by \textit{reverse-appropriating} these practices through the introduction of the new and formalized platform \cite{10.1145/3613905.3651034}. The co-opting of local infrastructure as a form of reverse appropriation is often framed within the discourse of \textit{technosolutionism}, which has been specifically identified as a mode of both data and digital colonialism \cite{mahmoudi2021race}. With the widespread adoption of platforms like WhatsApp and their pervasive presence in daily life, examining these platforms as infrastructures can provide valuable insights \cite{doi:10.1177/1461444816661553, 10.1145/3613905.3651034}. The study of infrastructures as \textit{sociotechnical geometries of power} \cite{graham2001splintering} within HCI has highlighted the significant inequalities and the persistent maximization of monopolistic behaviors \cite{10.1145/2702123.2702573, 10.1145/3491101.3505649, 10.1145/3557890}. 

I contribute to HCI research that highlights the importance of understanding how users who appropriate platforms (and infrastructure them) adapt to and identify challenges when the platforms change \cite{10.1145/3613904.3642590, 10.1145/3359175, doi:10.1177/1461444816629474}. By using the concept of \textit{infrastructuring}, I emphasize the ongoing effort where users, as key actors, dynamically and continually \textit{appropriate an infrastructuralized platform}. 


\subsection{Digital Colonialism and the Impact on Community-Based Informal Practices}

Digital colonialism refers to the imposition of digital norms and practices by powerful entities that reinforce existing power imbalances. It can be manifested through the control of digital infrastructures, data flows, and platforms that prioritize the interests of dominant actors \cite{mohamed2020decolonial}. This relates to informality and standards, as local, informal practices can be overshadowed or controlled by dominant standards, perpetuating these imbalances \cite{adamu2023no}. Colonial power structures often influence technology design by either providing universal solutions that overlook local specifics or assuming that people (especially from formerly colonized regions) cannot address their own issues \cite{10.1145/3544548.3581538, Couldry2021}-- thus often aiming for a global standard, emphasizing quantification and formalization \cite{lampland2009standards}, which frequently dominate reasoning frameworks \cite{pohawpatchoko2018cultural, cutajar2008knowledge}. HCI scholars have been notably critical of this trend in computing and design research. For example, Mignolo \cite{Mason2020} critiques global computerization driven by colonial expansion and calls on HCI scholars to challenge Eurocentric knowledge and value practices.

Within HCI, scholars have investigated trade practices in the Global South, focusing on how \textit{informality} facilitates and encourages unique business practices \cite{10.1145/3025453.3025643}. Additionally, it was noted that despite the prevalence of \textit{corporatized spaces}, local business places maintain distinct, localized characteristics that cater to the specific needs of communities \cite{10.1145/3025453.3025970}. As previously mentioned, researchers highlight that ritualized systems like WhatsApp are often preferred in informal practices over newer, more sophisticated platforms \cite{Cyr2008, 10.1145/3476057, 10.1145/3361116}. This preference is further influenced by the role of \textit{informal rumors}, which help individuals learn about new technologies and engage in collective sensemaking \cite{10.1145/3290605.3300563}. In this context, those who perceive themselves as marginalized often develop strategies to protect their interests while using services designed for more affluent users \cite{10.1145/3460112.3471961}. WhatsApp exemplifies this adaptation, showing how marginalized individuals appropriate it for their business \cite{10.1145/3613905.3651034}. 

Researchers have highlighted a common pattern where technologies initially adapted for local empowerment through context-specific modifications are later co-opted by large tech companies, leading to uneven power relations in design practices \cite{10.1145/3630106.3658934, 10.1145/3637316, 10.1145/1753326.1753522}- also termed repossession \cite{doi:10.1177/1461444816629474}. Additionally, \textit{material infrastructures and symbolic constructions}, such as the application itself and the ability of developers to design and redesign it in ways that maximize their profits, reinforce power imbalances and reflect colonial undertones \cite{doi:10.1177/1527476419831640}. In the past, Facebook's \textit{Free Basics} has been criticized as a form of digital colonialism, primarily because it was seen as a data collection tactic disguised as free internet access, while promoting large-scale technosolutionism that lacked substantial value \cite{Thorat2020}. Similarly, WhatsApp Business, Meta’s formal application, aims to enhance user engagement and sales but imposes costs on users in the Global South, who lack alternative platforms \cite{whatsapp_business, 10.1145/3613905.3651034}. Imposing on users with non-contextual designs pushes them toward Western business practices that emphasize formal competence, standardized procedures, impersonal task roles, and strong organizational commitment \cite{berger2020doing, Sinha1990}, which often contrast with traits like trust and empathy that are more prevalent in non-western contexts \cite{berger2020doing}. Researchers emphasize incorporating decolonial perspectives into appropriation to challenge Western standardization of knowledge and communication practices \cite{10.1145/3328020.3353927, wagner2020decolonizing}, while advocating for a reexamination of user-centered design approaches in non-normative cultures often overlooked by Western frameworks \cite{doi:10.1177/0162243910389594, 10.1145/2662155.2662195}.

 In conclusion, digital colonialism reinforces power imbalances by imposing standardized systems that overlook local needs, highlighting how small businesses in the Global South adapt informal practices to cope with disruptions and pushing users toward Westernized models. It emphasizes the need to recognize and respect local practices in design, challenging uniform digital norms. In this study, I address a critical gap in HCI by challenging \textit{top-down, paternalistic} views of technology in business contexts \cite{10.1145/3025453.3025643, cruz2021decolonizing}. My empirical work aims to demonstrate how users extend informal methods to manage business disruptions, especially when these disruptions push them toward standardized approaches that may not align with their sociotechnical and cultural contexts. By highlighting these adaptation practices, I contribute to HCI research on appropriation through informal practices \cite{10.1145/3328020.3353927, 10.1145/3571811}, underscoring the limitations of imposing uniform solutions and advocating for approaches that better respect and integrate the diverse needs of users.

%% file: methods.tex
\section{Methods}

I used a combination of purposeful and snowball sampling methods to recruit and interview 14 Indian WhatsApp users who use the platform to run their businesses.Consequently, I performed an inductive-- iterative thematic analysis of the data to address the research questions. The study and methods were approved by the review board. The specifics of this process are outlined below.

\subsection{Data Collection}

I used several different channels to recruit participants for my study, as they were part of a very informal network of workers, and accessing them through more formal communication methods was challenging \cite{agarwala2009economic, 10.1145/3025453.3025643, 10.1145/3290605.3300563}. Therefore, I employed purposeful \cite{suri2011purposeful} and snowball sampling \cite{parker2019snowball} across four platforms: WhatsApp, Facebook, Instagram, and Quora. The recruitment messages varied depending on the context of each platform. On WhatsApp, I started with personal contacts (n=12) who were involved in business and whom I had previously interacted with as a customer. I asked them to refer individuals who might be interested in my study and to contact me via my personal WhatsApp, ensuring a degree of separation between myself and the participants. A total of 16 people were referred, among which 7 responded when I messaged them and agreed to do the interview. For Facebook, I joined three public groups related to WhatsApp Business in India, identified through keywords such as "WhatsApp Business in India," "Indian Businesses using WhatsApp," and "WhatsApp for Business in India." I engaged with posts advertising WhatsApp businesses and used links provided to contact WhatsApp group admins (n=42) among which 3 responded, and agreed to interview for the study. On Instagram, I targeted accounts that promoted local Indian businesses and reached out via direct message to the admins (n=17) of businesses that had a WhatsApp option listed on their profiles, and 2 of the admins responded, and interviewed with me using this method. Lastly, on Quora, I posted a recruitment message in three channels related to WhatsApp Business in India and used similar keywords to those used on Facebook to invite interested individuals to contact me via email, and 2 participants were interviewed via this method. As potential participants contacted me via WhatsApp or email, I screened them to ensure their home-based businesses met the legal turnover range and local criteria for small businesses in India (non--public company with share capital below Rs. 40 million.) \cite{govindia2023}. Recruitment ceased when data saturation was reached, meaning no new insights emerged from interviews \cite{Charmaz2012}. This was confirmed through corresponding preliminary analysis to ensure all research questions were thoroughly addressed.

 \begin{table*}[t]
\footnotesize
  \caption{Self-Reported Participant Demographics, where an asterisk (*) indicates that the business was operated on a part-time basis, while the absence of an asterisk signifies full-time involvement. Despite WhatsApp's payment options, all participants used Unified Payments Interface (UPI), Google Pay, or cash.}
  \label{tab:freq}
\resizebox{\textwidth}{!}{%
\begin{tabular}{p{1cm} p{1cm} p{1cm} p{1cm} p{2cm} p{1.7cm} p{2.7cm}}  
   \toprule    ID&Age&Gender&Education&Monthly Income& Approximate number of Customers&Type of Business\\
    \midrule
    
    P1 & 32& Female& High School & Rs. 5000-8000 & 500&Local handmade sarees and kurtas (Reselling) \\
    P2 & 28& Female& College diploma & Rs. 1500- 1200 & 230&Food Business* \\
    P3 & 31& Female& High School & Rs. 15000- 20000 & 2000& Handmade dress materials*\\
    P4 & 43& Female& High School & Rs. 8000-10000 & 330& Food Business*\\
    P5 & 48& Female& High School & Rs. 9000-12000 & 5000& Beauty Services*\\ 
    P6 & 51& Male& College Diploma & Rs. 9000-12000 & 8000&Reselling dress materials and sarees \\
    P7 & 33& Female& College Diploma & Rs. 800-1200 & 700&Food Business*\\
    P8 & 31& Female& College Diploma & Rs. 1000-3000 & 120&Salon services* \\
    P9 & 37& Female& College Diploma & Rs. 2000-4000 & 100& Handmade Artwork\\
    P10 & 35& Female& College Diploma & Rs. 3000-5000 & 170& Handmade Artwork*\\
    P11 & 39& Female& High School & Rs. 500-1000 & 40&Food Business*\\
    P12 & 48& Female& High School & Rs. 1000-1500 & 300& Reselling locally manufactured sarees\\
    P13 & 52& Male& High School & Rs. 8000-10000 & 200&Reselling \\
    P14 & 54& Female& High School & Rs. 10000-15000 & 6000&Makeup and Salon \\
\bottomrule
\end{tabular}%
}
\end{table*}

\subsection{Participants and Interviews}

I conducted semi-structured interviews with 14 participants in India who had spent between two and nine years running home-based small businesses on WhatsApp. Five participants were involved in the clothing business, purchasing garments from artisans or manufacturers to sell via WhatsApp. Four participants sold food through WhatsApp for pickup or delivery. Three worked in the beauty industry, offering home services or selling courses and products. Finally, two ran creative businesses, making and selling mud, clay, or painted artifacts. Participants had varied amounts of digital skills, and in particular, this manifested throughout the interviews, with many coming from cultures where they did not receive much support or access to digital resources. For some, the lack of formal training or mentorship in digital technologies meant they had to rely heavily on self-learning, often navigating challenges alone. Others highlighted the impact of cultural norms that did not prioritize or encourage digital literacy, which further shaped their engagement with technology. These differences in background and support systems contributed to distinct approaches and levels of comfort when using digital tools, with some participants feeling more confident and others struggling to keep up in an increasingly digital world. Most (n=13) started by using WhatsApp’s personal app and either continued using it for business or used both the personal and business versions of the app. Their reported customer numbers ranged from 40 to 8,000. Additional participant demographics were collected during the interviews and are listed in table \ref{tab:freq}.

The interview protocol was semi-structured and participant-centered, designed to foster open dialogue and make participants feel comfortable. It began with general questions about their work, businesses, and use of WhatsApp, allowing for a broad understanding of their context. To address the first research question, the protocol then explored the transition from WhatsApp as a personal communication tool to its use in business, focusing on how participants integrate the platform into their professional lives. Questions probed the technical aspects of using WhatsApp in business, including its benefits, challenges, and role in supporting business activities, aiming to understand its evolving use in both personal and professional spheres. For instance, participants were asked: "How did you first start using WhatsApp for business purposes?" and "What features of WhatsApp do you find most useful in your business?" The interview also focused on distinguishing between WhatsApp (the personal app) and WhatsApp Business, with questions like "What made you choose to use WhatsApp Business instead of the regular WhatsApp app?" or "What are the main differences you’ve noticed between WhatsApp and WhatsApp Business in your day-to-day operations?" This allowed for insights into participants' preferences, and challenges in using either application. Following this, the interview shifted to understanding the process of adaptation, with questions aimed at uncovering how recent and anticipated changes in WhatsApp usage were perceived. For example, interviewees were asked, "Have you noticed any significant changes in how you use WhatsApp for business in the past year?" and "What changes do you foresee in the future, and how do you feel about them?" In many cases, participants discussed these shifts spontaneously, providing rich, unprompted insights into their experiences and expectations. As the sole author, I also memoed my experiences during and immediately  after each interview, which served as a way to guide the analysis, and motivate emerging themes \cite{10.1145/3359174, 10.1145/3491101.3516392}. The semi-structured interview guide is provided as an artifact, though it is brief and broad, as is typical in interviews where questions evolve and depart from the protocol as the conversation unfolds \cite{roulston2018qualitative}.

All interviews were conducted via WhatsApp calls, and Zoom was utilized to record them using a loudspeaker. Additionally, each interview began by informing the participant of their rights and obtaining their consent to be recorded. The interviews, conducted in Hindi, English, or Bengali, lasted between 15 and 120 minutes. One interview lasted only 15 minutes due to the participant's limited access to Wi-Fi and inability to reschedule because of personal issues. The average interview duration was 37 minutes, with a standard deviation of 27.47 minutes and a median duration of 40 minutes. I translated and transcribed the interviews directly into English for analysis. 

\subsection{Data Analysis and Positionality}

As data were being collected, I simultaneously analyzed the existing transcripts \cite{smith2024qualitative}. This approach, inspired by grounded theory \cite{charmaz2015grounded}, allowed me to refine and adjust the interview protocol based on preliminary findings and to enhance the ongoing analysis. Although inspired by grounded theory, I did not strictly adhere to this methodology. Instead, I incorporated select elements of grounded theory to guide the development of codes and facilitate the thematic analysis \cite{braun2012thematic}.

Analysis began with a thorough review of the transcripts, during which open codes \cite{strauss2004open} were assigned at a descriptive level (for eg., communication block, technical difficulty, business tactics) \cite{terry2017thematic}. After approximately three rounds of open coding, a more focused coding \cite{thornberg2014grounded} approach was employed to identify relevant themes (for eg., infrastructuring as ongoing labor ) that addressed the research questions more directly. The reported findings show an even higher level of analysis, as more codes were conflated and corresponding themes for the research questions emerged. This research adopts an interpretivist approach \cite{10.1145/3633200}, wherein themes and codes are subject to interpretation \cite{10.1145/3359174}. As such, there is no singular, fixed understanding of these themes and codes. This perspective challenges the notion of objective meaning, emphasizing that my subjectivity, positionality, and interpretive stance contribute to deriving meaning from the data \cite{10.1145/3359174}. To motivate this interpretive analysis, memos were triangulated with the transcripts to produce the insights and findings presented in the following sections \cite{10.1145/3359174}. The final codebook is shared as an artifact.

As a researcher born and raised in India with personal connections to individuals who use WhatsApp for business, my understanding of their socioeconomic backgrounds, views, beliefs, and motivations is influenced by my own experiences and relationships. In this research, I made a deliberate effort to identify and address my own subjectivity during the analysis of data by being explicit in allowing the participant to guide the conversations, through semi-structured interviews \cite{berger2015now, doi:10.1177/1468794112439005}. I also took into account the power dynamics when interviewing individuals who might be wary of my current affiliation with a university in the United States. To address these concerns, I established at the outset that the interviews would be participant-led and semi-structured, drawing on feminist research methodologies that emphasize the empowerment and agency of participants. This approach has been used to engage traditionally marginalized users \cite{doi:10.1177/1468794112439005}. While I had key questions aligned with the research objectives, I allowed participants to guide the conversation and raise emerging concerns. This not only gave them more autonomy but also facilitated the collection of more authentic information, encouraging participants to express their perspectives and experiences more freely \cite{doi:10.1177/1468794112439005}. Acknowledging the complex nature of the \textit{"at home"} versus \textit{"not at home"} discourse, I focused on identifying commonalities and embraced the absence of such binaries as a way to foster honesty and genuine respect \cite{doi:10.1177/1468794114550440}. This approach was crucial, given that there were many aspects I did not know or experience because of my different affiliation. Thus, while my relative positionality influenced the analysis, I employed rigorous inductive coding to ensure that the data itself guided the findings.

%% file: findings.tex
\section{Findings}

The findings, divided into three sections, address the corresponding research questions. The first section explains how users adapt business practices on WhatsApp, while the next two sections explore the implications of ongoing infrastructuring on users, answering the second question.

\subsection{Translating and Enacting the Personal to the Professional}

In this section, I discuss the interviewees' experiences of using WhatsApp for business, focusing on how they came to adopt the platform. While WhatsApp’s informal and personal nature played a key role in its use, it also blended with professional tasks, creating additional labor that was often overlooked. This led business owners to have dynamic, evolving views on the role WhatsApp played in their work.

\subsubsection{Personal Familiarity and Ease of Transition}

Interviewees explained that their adoption of WhatsApp for Business was a natural extension of personal use, often driven by job losses, family responsibilities, the pandemic, or changes in family members' professional circumstances. For example, P5 appreciated the convenience of WhatsApp as a system for managing her business, reflecting on how her professional work integrated seamlessly with her personal use: "\textit{WhatsApp is convenient. It’s easier to use than other platforms, and it’s where most of my customers are.}" Here, P5 described how using WhatsApp as a natural platform was not only convenient for her but also for her customers, who were already on WhatsApp and did not have to make an extra effort to engage with her business. Additionally, P1 emphasized that, given her existing familiarity with the app, it was a "\textit{smart move to leverage it for a wider range of purposes}". P9 shared a similar perspective, noting that WhatsApp felt like a natural progression due to its familiarity. She explained that using WhatsApp casually for personal connections made it \textit{"the most straightforward option"} when everything shifted to virtual platforms. According to P9, \textit{"It was like a lifeline—I could instantly reach my customers, share updates, and manage orders all from an app I was already comfortable with."} She also emphasized that since everyone was already on WhatsApp, it \textit{"avoided the need to transition to a new platform."} P12 confirmed this choice of WhatsApp for her business, explaining that she felt familiar with the platform and its nuances, or at least with the aspects necessary to run her business effectively-

\begin{quote}
     \textit{I’d already been using WhatsApp to make groups, video call my friends and share messages and interesting information, so it was very obvious to use it for my business. I started using it to chat with customers, update them on new stuff, and even handle orders.}
\end{quote}

In particular, P1 added that her WhatsApp business was a direct response to the challenges posed by the pandemic: "\textit{I started using WhatsApp for business about 5 or 6 years ago, but the real shift happened during the pandemic when I had to move everything online due to shop closures.}" This sentiment was echoed by P3, who said,\textit{“I used to run my boutique offline. But when everything closed, I had to find a way to keep the business going. WhatsApp allowed me to continue reaching my customers and even expand my network.}” Given that the customers and staff of the boutique were already familiar and somewhat close, WhatsApp felt like a natural extension of their existing communication, even though people were not accustomed to \textit{app-based businesses} (P11). It simply served as an intuitive tool to stay connected and continue business as usual.

\subsubsection{Influence of Personal Connections and Informal Engagement}

Some interviewees mentioned that a personal connection had introduced or encouraged them to use WhatsApp for business. As P2 explained: \textit{"My sister started using WhatsApp for business during lockdown. I saw that she could do it so easily, so why not try it? Since I was already familiar with the app, it felt pretty easy to shift to!"} This guidance from someone within their personal circle, also provided confidence. 

Many individuals also aspired to use WhatsApp to extend their personal use into a side project, and referred to their business as a \textit{hobby} that even their "\textit{ close family, like [their] my parents, don't know about}" (P7). P7 shared, 

\begin{quote}
    \textit{I don’t really consider myself a business person— and even my close family like my parents don’t know much about it. WhatsApp is just really convenient for me to handle these little orders and keep in touch with everyone without making a big fuss about it. Now I have more than 300 daily customers but its just my everyday thing, I would cook even if I did not sell food so its just like doing my regular thing.} 
\end{quote}

Some interviewees framed that their engagement with WhatsApp was informal. For example, P7, was adamant about labeling her activities as informal engagements rather than a business, often referring to them as hobbies or something "\textit{on the side}". Additionally, P12 explicitly stated that these pursuits were outside the construct of any type of "\textit{serious endeavour}" and P11 added, they did not require "\textit{professional expertise}" (P11). P4 exemplified this by saying that her WhatsApp business activities were personal ventures, explicitly denying its existence as a formal business. She explained, how her customers are "\textit{friends or friends of friends, and because they know [her] me personally, it almost feels like [she] I just cook(s) for [her] my friends}" while at the same time admitting that she "\textit{benefits financially, and helps support [her] my home}". This reflected how participants often described their WhatsApp business activities as informal engagements, where the personal nature of their interactions with customers made it difficult to view their work as a formal business. However, this view was dynamic, nuanced and also debated by some, as explained in the next section.

\subsubsection{Labor, Legitimacy, and Recognition of Professionalism}

The prevailing narrative about WhatsApp had largely depicted it as a tool for personal communication, overshadowing its potential and significance in business contexts, as illustrated by the participants. However, P14, who owned a salon during the pandemic but had to shut down her business, was open about how her now WhatsApp business was valid and required significant planning, infrastructuring, and management. P14 detailed the effort involved in tasks such as "\textit{scheduling messages, sharing rates, and ensuring the right products or content is sent to the right people}." She explained, 

\begin{quote}
   \textit{You know, this business takes up so much of my time. I’m not sure if I want to keep doing it long-term because I almost make as much as my husband does [as a middle school teacher]. Sometimes I feel like I’m neglecting cooking and taking care of the house. People think because it’s a WhatsApp business that it’s just chatting and videos, but they don’t see the effort I put into building relationships with my customers. It’s only when the money comes in, like during wedding season, that people realize I’m actually running a serious business.}
\end{quote}

Here, the participant was concerned with the frustration of having a business that others dismissed as insignificant. She felt disheartened that its value was only recognized when it benefited the family. Additionally, she was unhappy that the affective and relational labor she invested in building connections and trust within the business was not acknowledged, making the business seem less legitimate than she felt it was.

However, another participant, P2, echoed a different view, suggesting that WhatsApp Business was relatively informal and this informality was what made her comfortable. She explained, \textit{“[WhatsApp] just has all these contacts, and I can add them to my business group to sell things I would usually sell”} and \textit{"It's nothing extra I do, but I can make some extra money, and it's helpful like that"}-- specifying how she herself did not see any of this as extranous labor, aligned with feminist interpretations of labor \cite{lokot2020unequal}.

These perspectives illustrate how both P2 and P14 defined and developed their understandings of \textit{labor}—in deciding on the types of labor that were considered legitimate and worthy of recognition in a business context. Despite the \textit{immaterial labors} P2 invested in, she felt that these efforts were not \textit{“serious enough”}, even though they contributed to material gains. In this way, there was a range of perspectives that defied simple binaries in how participants viewed their own businesses and their use of WhatsApp. P8 added complexity to it, as she navigated a middle ground between running a full-fledged business and pursuing a hobby project, reflecting a variety of personal approaches and comfort levels sharing- 

\begin{quote}
    \textit{I offer part-time home salon services, especially in high demand during wedding season. At other times, my earnings are minimal, so it’s not very formal. I just understand what people want, and since I used to do my friends' makeup, I thought, why not turn this into a service and charge a bit for it?}
\end{quote}

In conclusion, WhatsApp produced diverse experiences around labor, legitimacy, and professionalism, shaped by users' own statuses, situated ideas of labor, and their capabilities. These varying experiences also reflected how WhatsApp allowed users to navigate and negotiate labor and professionalism, informed by their personal circumstances and unique perspectives on what counted as legitimate and professional-- and this flexibility was useful to them. 

Next, I highlight how WhatsApp's evolution required users to adapt without explicit compensation or recognition as Meta engaged in coercive professionalization, during which this flexibility was lost. Users often navigated through workarounds; however, as further explained, these adjustments were frequently unacknowledged by the users themselves.

\subsection{Continual Infrastructuring as a way of Use}
 
In this section, I examine how users adapted to platform changes by learning through experience, often with the support of personal networks, and the time investment required in the process.

\subsubsection{Learning by Experience}

Despite the obstacles presented by frequent interface updates, users developed methods to adapt and integrate new features into their routines, treating each change as an opportunity to refine their use of the app. For instance, P10 exemplified,

\begin{quote}
\textit{Whenever I was chatting with my sister and noticed a new button or feature, I would try to use it right away. This way, when I needed to use it for my business, I already knew how it worked.}
\end{quote}

This underscored how users’ extensive personal use of WhatsApp allowed them to learn and adapt to its features beyond the context of business. Even those who might not have considered themselves technologically adept assumed WhatsApp to be "\textit{easier, with little need for external manuals}" (P5). P9 additionally explained- 

\begin{quote}
    \textit{You know I don't understand Facebook and Instagram when they change. But I think WhatsApp is just to message-- and even if there is more now, I try to always only stick to messaging. That way, if I know how to send one message I know how to use it well- and sometimes that may take me hours and hours of trying to understand the new change, but once I do, I am back at it.}
\end{quote}

Through this process, users leveraged their routine interactions with the app to develop and refine their skills. By experimenting with new features in personal messaging, users effectively navigated the learning curve in a low-risk environment before applying these skills in business settings. This method of experiential learning demonstrated that frequent engagement with the app, as a routine infrastructure, facilitated a deeper understanding of its features, despite ongoing updates and changes.

However, this notion of continual change was also approached with caution. Both P10 and P9 expressed concerns that if the app continued to evolve with \textit{too many features} (P9), it might expand beyond its primary function of messaging. Participants speculated about the potential challenges this \textit{complication} (P15) could pose and expressed uncertainty about how to manage such changes. For instance, P10 shared,

\begin{quote}
    \textit{I feel like WhatsApp is turning into something like Instagram, which my daughter only knows how to use. I need WhatsApp to stay simple so I can keep using it. If it adds too many features and becomes too complicated, I won’t see the point, and neither will my friends or customers.}
\end{quote}

In order to continue, as P4 shared- "\textit{figuring out WhatsApp, because I [she] will anyways use it for chatting but then when it becomes too complex and I [she] am [is] no longer understanding}", explaining if the complexity of the tool were to increase to a point that it becomes "\textit{unrecognizable and not just a text message app anymore}", then "\textit{[she] would not have the independence to continue doing this kind of business or even continue using it easily enough}". While learning through experience made WhatsApp a preferred tool, participants also noted its limitations and discussed its boundaries. They explained that relying on routine infrastructuring was effective only as long as those experiences did not become overly cumbersome or obstructive to new ways of adapting to the platform.

\subsubsection{Personalized Networked Learning}

Participants explained how they used their personal connections on WhatsApp to manage changes, even when they did not fully understand all of them. The core functionality of the tool inherently supported this form of learning by enabling users to apply the informal, personal communication aspects of the tool to their professional contexts, thereby facilitating their adaptation and learning.

Using her personal network, P3 shared her experience with infrastructuring only after confirming with her social connections. She explained how she had friends in her community with whom she was accustomed to \textit{discussing} issues, and how this process continued iteratively as new changes emerged - involving the same group of people to discuss and develop solutions.


Consequently, others had other set people in their personal lives who also used WhatsApp similarly, working collaboratively through a trial and error method. P8 shared,

\begin{quote}
    \textit{I started using WhatsApp with help from my sister. We learned together how to post updates on my status and in groups. [...] Whenever either of us finds something useful, we share it with each other.}
\end{quote}

Finally, some participants reflected on their personal use of WhatsApp to explain how collaborative learning unfolded. As \textit{P13} shared, \textit{"It feels like there are always changes happening online. I usually reach out to my friends on WhatsApp, and we end up discussing things and figuring stuff out together, even if it’s by mistake."}

Here, the personal nature of WhatsApp played a pivotal role in enhancing the collaborative learning process. The platform's accessibility and informal communication style enabled participants to engage in spontaneous discussions, which, even if unintentional, led to shared problem-solving and learning. This ease of connection allowed for a flexible and dynamic approach to learning, where adjustments were made naturally in response to ongoing interactions. Thus, WhatsApp facilitated not only communication but also collaborative adjustment and discovery in an organic way.

However, not everyone had the ability to immediately access networks that could provide intellectual or practical support within their immediate surroundings and social circles. For instance, P6 who was frustrated with the interface changes, shared- "\textit{I know the inner features are the same [..] but sometimes the status icon moves from the top to the bottom and back again. It’s tough to keep up.}" 

Similarly, P1’s difficulty with disappearing interface elements underscored how even simple tasks become daunting when users can’t rely on a stable, predictable design:

\begin{quote} \textit{Sometimes, I’ll see the three dots at the top of the screen, and then they just disappear. A few days later, they’re back again. It’s frustrating because it makes simple tasks, like choosing who to message, much more complicated.} \end{quote}

Her reliance on her daughter for guidance further highlighted her difficulty accessing external support—\textit{‘My daughter understands how it works, but she's very busy and living abroad’}—pointing to the limitations of informal help and the isolation that can result when support isn't immediately available.

Overall, for P1, the workflow and changes were not intuitive. Although she tried to use her network, she was less fortunate than others because her contacts were not in close proximity. As a result, while personalized learning from social contacts was available, it was not as convenient or straightforward for her.

Overall, these accounts suggested that frustration stemmed not only from navigating changing interfaces but also from the ongoing reliance on external help and the limited availability of immediate, local support, highlighting the challenges users faced in adapting to both the evolving technology and the scarcity of accessible, reliable resources. However, among the available workarounds, infrastructuring through personal networks and informal communication was the most useful and sought-after method for learning. Participants used WhatsApp’s core messaging features to navigate time-consuming changes, with some benefiting from collaborative efforts within their social circles, while others faced challenges due to limited access to supportive networks. These experiences highlight how infrastructuring, shaped by personal and social contexts, was crucial in managing technological changes. The next section discusses how this process often became laborious and time-consuming.

\subsubsection{Time-Consuming Adaptation}

The process of adapting to WhatsApp’s frequent updates was both time-consuming and overwhelming. Infrastructuring, characterized by its continuous nature and constant need for learning, further compounded the time investment required at each step of the adaptation process. P7 shared:

\begin{quote}
\textit{It takes a lot of time to learn how to deal with WhatsApp’s changes- and I already tell my husband and son that this is something on the side and when I spend this much time, and I still end up not really doing as much business because I am still understanding the changes it becomes pointless!}
\end{quote}

This quote highlighted the considerable effort required to stay abreast of changes while managing the demands of  other responsibilities. P4 additionally operationalized a change and how it impacted her, 

\begin{quote}
    \textit{I keep wasting time trying to figure out these new changes. One day, I could send my product to all my groups, and then suddenly I could only send it to five. It took me a whole week to understand the new update, and by then, my clients had moved on to other stores or they wanted other things. A lot of them were confused about why I couldn’t get products to them properly or reply on time.}
\end{quote}

Both P7 and P4 interpret and perceive their work within their unique sociocultural contexts, and the ongoing process of infrastructuring consumed more time than they had available. Similarly, the ability to allocate time, or the lack thereof, further influenced what it meant to successfully appropriate the platform through infrastructuring. As another participant, P3, explained, she didn't mind \textit{discussing with her friend group} how the ways in which WhatsApp was changing could be "\textit{a hit or a miss for [her] business}". She elaborated that the changes were manageable for her, as she was a "\textit{recent retiree from a bank job}" and understood why companies needed to implement these updates but felt that they missed the point. She explained-

\begin{quote}
    \textit{I get that WhatsApp is trying to cut down on spam, but it doesn't really help small businesses like ours. We still get spam calls and video calls. We just share these numbers in our friends' group and block them ourselves.}
\end{quote}

Others, like P13, found the changes disruptive and demotivating. P13 explained that he \textit{randomly received video calls from unknown numbers}, and found it difficult to block them. He said, "\textit{All these new symbols and buttons are overwhelming, and now they're asking me if I want some automated process. No, I just want to block the spam!}", adding that "\textit{I don't have so much time to accost WhatsApp's changes and also keep up with all the nonsense}".

Likewise, both P2 and P4 were extremely dissatisfied as due to their personal contexts, their learning curves were quite steep. P4 mentioned, \textit{"You know, now that it’s so overwhelming, I can’t continue working as domestic help. I have to choose between quitting everything and focusing on this business or finding time to work out. It’s always about having to choose—whether to stay at home and do this or to work outside."} P2 shared a similar experience, explaining that "\textit{her ability to keep up with changes was interfering with her household responsibilities}" and her husband might "\textit{soon ask her to stop.}"

In conclusion, adapting to WhatsApp’s frequent updates highlighted the significant role of infrastructuring in users’ experiences. The continual need to manage these updates—a core aspect of infrastructuring—was both time-consuming and overwhelming for many. Participants such as P7 and P4 noted that keeping up with these changes often consumed more time than they had available, impacting their ability to manage their businesses and other responsibilities effectively. This situation revealed that the ability to cope with updates was influenced by life factors, with those having stable personal lives and strong social support handling updates better, while others facing personal challenges or lacking support experienced more difficulties. It is also of note, that while WhatsApp's informal nature initially facilitated a seamless transition from personal to professional use, it has evolved into a more structured and demanding platform. This evolution has led to a form of coercive professionalization, where users, particularly marginalized entrepreneurs, are compelled to adapt to constant and \textit{standard} platform changes which do not support their local needs. I discuss more about coercive professionalization in the discussions.

\subsection{Normalizing Invisibilized labors of Continual Infrastructuring}

In this section, I share how participants demonstrated a resigned acceptance of WhatsApp's continual changes, normalizing the process of learning and adapting as an inherent part of their daily lives. This acceptance extended beyond their businesses, reflecting a broader pattern where adjusting to challenges and navigating constant updates with technologies became a routine aspect of their labor, shaped by their broader experiences of marginalization and the necessity of making do with available resources.

\subsubsection{Continual Appropriation to Infrastructure through Changes}

As WhatsApp continually changed and redesigned its user interface and functionalities, the participants I interviewed frequently encountered issues and had to figure out workarounds. However, when asked about this ongoing challenge, they often maintained that they did not view it as extra work. They accepted these difficulties as normal, given their status as \textit{small players} who \textit{always had to find workarounds}, since the platform was not designed with their needs in mind. P10 explicated this-

\begin{quote}
    \textit{WhatsApp keeps changing so much—now you need verification, your number can get blocked, and there are communities instead of just groups-- its all so professional [..] I’m not very tech-savvy, so all these updates are really tough for me. I don’t want to complain too much because I’m just glad to still have WhatsApp. [..] I can’t switch to other platforms; I wouldn’t understand them, and neither would my customers.}
\end{quote}

The concern extended beyond personal inconvenience, presenting the participant with issues related to customer retention. If WhatsApp became unusable or if users were required to switch platforms, they faced the dual challenge of adapting to a new, unfamiliar system while also managing the risk of losing their existing customer base. Users felt \textit{trapped in this} (P14) situation where they had to endure ongoing changes to maintain their business connections, as any switch could disrupt their operations and alienate their customers. Likewise, others such as P3 and P8 shared their sentiment of being extremely careful in terms of their expectations, with P8 sharing- "\textit{I try not to anticipate too much because the platform is so unpredictable now. It feels like every time I get used to something, it’s replaced with something new}". P3 shared this opinion, saying- \textit{I don't try anything new, I only use what I need}- particularly due to prior experiences of attempting to set up an infrastructure on something new, only to have that made paid and \textit{too expensive to use}. However, both P3 and P8 also agreed with the others, in finishing \textit{But you know what? At least WhatsApp is still here. I can adjust my strategy as needed, but at the end of the day, it is WhatsApp. Even if they charge me a bit, I know I won't find anything else like it.} (P8). P3 added to this-

\begin{quote}
    \textit{My husband and daughter don’t have time to help me. It’s sad if WhatsApp increases its price or asks me to make websites or something, but I’ll have to pay it. If I can’t, I’ll have to stop my business because I can’t do it anywhere else.}
\end{quote}

As an example of the challenges posed by platform complexity, P1 noted that WhatsApp's increasing sophistication could make businesses appear more legitimate, but due to her sociocultural constraints, it might force her to shut down her business.

\begin{quote}
    \textit{If the platform became more complex and made my business look more legitimate with verification and everything, it might mean letting everyone in my extended family know about it. This visibility could lead to me having to close the business, as my family might not support it.}
\end{quote}

To manage these updates, P1 tried to keep her business operations unchanged unless WhatsApp made automatic adjustments. However, she often found that any action she took seemed to trigger further changes, as she described: \textit{"It’s like anything I touch, some change takes place."} To cope with this, she often shut down her business for long periods of time. 

\subsubsection{Resigned Acceptance of Normalized Labor}

More often than not, unnoticed effort highlighted the constant work needed to incorporate new features into daily business practices-- these were often dismissed by the participants. Instead, they normalized the extraneous labor of constant infrastructuring as something implicit and expected of them, as P9 shared- "\textit{Why will companies or anyone make anything for us? We are poor people}". Like P9, many participants, who faced marginalization through various personal identities, viewed WhatsApp’s availability as a significant benefit that they were undeserving of. P11, who described her WhatsApp business use as a way to \textit{make a little more money so [they] can eat outside once in a while}, further shared "\textit{But now; Business on WhatsApp is tough now—but am I really a businesswoman?}".

This perspective revealed that business owners were generally satisfied with having access to WhatsApp, accepting its evolving features without scrutinizing how their use was monetized or exploited. For example, P7 explained,

\begin{quote}
\textit{I've been using WhatsApp for my business for many years now. Even though WhatsApp keeps changing and updating, I'm just glad to have it. I manage with whatever new features come along and make it work for me. For a small home business, I don't need all the fancy-professional tools that big companies might use. I take things as they come and handle any changes because, honestly, having WhatsApp is a huge help. I realize that some of the new tools might be designed more for bigger companies, that need formal presence but I make do with what I have.}
\end{quote}

Here, the participant thought that she lacked agency to assert that the platform was designed with her needs in mind. Instead, she referred to \textit{big companies}, effectively downplaying her own needs as a businesswoman and questioning her own entitlement to the platform's features-- another description of the professionalized platform that she was coerced to adapt to. Inspite of the experienced marginalization, they remained appreciative. P12, whose son resided in the United States, shared her thoughts on this situation, stating:

\begin{quote}
    \textit{My son in the US tells me that WhatsApp Business is going to be a big deal, but I’m not so sure. I’ve been using WhatsApp for Business for the past nine years and never thought much of it. Now that it’s more popular, suddenly my account gets blocked, my numbers are reaching too many people, and many of the new features seem to come with a cost. I don’t have the money for that, but I guess I’m just older now, so I’ll just accept whatever comes my way}
\end{quote}

Here, the normalized acceptance of \textit{whatever comes} reflects a sentiment where participants felt they were undeserving of WhatsApp amid its changes and developments. This idea was intertwined with their identities, where they were used to experiencing marginalization, as P5 explained, 

\begin{quote}
    \textit{I feel like I’m not really cut out for doing business, and thats what my son and husband keep saying [..] All this complex technology confuses me—I only understand and use WhatsApp. No matter what happens, I’m just so happy it’s here [..] I’ll be grateful for whatever I’ve had}
\end{quote}

They incorporated a sense of undeservedness into their understanding of the platform's continual modifications, which diminished self-confidence, while rationalizing that enduring these challenges was preferable to seeking alternatives due to gratitude for the service's availability. This prevented them from critically evaluating or negotiating the platform's conditions, especially since Meta obscured how it benefited from the arrangement and appropriated the infrastructure users built. As a result, users faced the pressures of adapting to constant updates and managing potential cost increases, while understanding that without WhatsApp, their businesses would face significant operational difficulties.

%% file: discussions.tex
\section{Discussion}

The findings reveal that WhatsApp’s rapid adoption for business purposes marks a significant shift, as personal communication tools are appropriated for professional use within complex sociotechnical contexts. This aligns with prior research showing how users extend the platform’s functionality to support their professional activities \cite{rupvcic2020emergence, dominguez2020distributed, 10.1145/3025453.3025643}. As WhatsApp Business professionalized, Meta engaged in reverse appropriation by monetizing the platform and altering its infrastructure \cite{10.1145/3613905.3651034, owen2024monetization}. This led to continual infrastructuring, where users had to frequently adapt their practices. Despite their efforts, users often undervalued this labor and invisibilized their own legitimate practices \cite{crain2016invisible, doi:10.1177/1050651920979999}. Understanding users' practices in why they perceive their own efforts as unworthy is crucial for understanding the broader implications of reverse appropriation and infrastructuring on their experiences \cite{munro2012social, 10.1145/3555584, 10.1145/3415170}.

\subsection{Appropriation through Personal Use}

\label{appper}

The \textit{personal} use and ritualization of WhatsApp \cite{doi:10.1177/01968599221095177} allow individuals to interact with and adapt the platform in informal ways. This flexibility helps them achieve more than they might with more rigid, formal tools. For example, users like P8 and P3 use WhatsApp’s personal communication features to indulge in \textit{informal rumors} \cite{10.1145/3290605.3300563} to collaboratively address business challenges and adapt to new updates. This community-driven approach helps them manage their businesses effectively, even with limited digital knowledge and experience \cite{10.1145/3637429}. Through this adaptation, users have appropriated WhatsApp to meet their needs, benefiting both themselves and Meta \cite{10.1145/3313831.3376201}. The platform’s simplicity and personal nature are crucial for its success in these informal business contexts. A more complex and professional tool might not fit well with the users’ sociotechnical environments, potentially overwhelming or deterring them from pursuing entrepreneurial opportunities.


When users are required to adapt to more formal structures, the situation changes significantly. As WhatsApp evolves and potentially introduces more complex features, users who are accustomed to its informal, personal use struggle with the demands of the professional setting \cite{10.1145/3613905.3651034}. This creates a tension: while the personal nature of tools like WhatsApp promotes innovation and accessibility, their evolution to meet professional needs must be carefully managed to avoid alienating users who value their simplicity and familiarity. Thus, I advocate for further research on artifacts that are inherently personal, appropriated, and ritualized to be used for a wider range of tasks. Examples of this can be shared points of internet connectivity like cyber cafes, telehealth infrastructures, and entrepreneurial applications, among others. Some of these have already revealed design \cite{furuholt2018role} and sociotechnical implications \cite{Rangaswamy2011CuttingChai}, which could be leveraged and co-opted to \textit{center} marginalized populations to legitimize and solidify their appropriation practices \cite{berger2020doing}. In the next section, I explore how users have adapted to the changes.

\subsection{Coercive Professionalization and its Consequences}

In this section, I examine the experiences of small business owners adapting to WhatsApp's changes through the concepts of appropriation \cite{10.1145/3613904.3642590}, colonization \cite{doi:10.1177/1527476419831640}, and infrastructuring \cite{doi:10.1177/00027649921955326}, with a focus on their sense of undeservedness. This study and previous research show that WhatsApp has reached a state of \textit{infrastructuralization} through unique processes of \textit{appropriation} \cite{10.1145/3613905.3651034} and \textit{ritualization} \cite{doi:10.1177/01968599221095177}. Users have extended the app’s personal use into broader, non-personal contexts. However, WhatsApp’s evolving features and monetization strategies impose a standardized, professional framework that is perceived as mainly benefiting larger businesses and marginalizing small business owners. For example, P1 expressed concern that WhatsApp’s push towards more \textit{legitimizing} features might compel her business into an unwanted level of visibility-- a direct consequence of the ways in which WhatsApp was wanting its users to professionalize. Likewise, P9 and P10 were concerned that if the app continued to evolve with more features, it might expand beyond its primary messaging function, potentially forcing them to use more professional tools. Others speculated about the complications this could cause and were uncertain about how to manage such changes, fearing it might push the platform into a realm they were uncomfortable with. This is what I note as perceived \textit{coercive professionalization}, where users are pressured to \textbf{adhere to global, often Eurocentric, professional standards, often ignoring their specific local contexts}. Similar to a colonial-style reverse-appropriation \cite{10.1145/3613904.3642590}, this process forces users to conform to norms that do not reflect their unique needs or cultural situations-- impacting how users perceive and respond to the platform’s changes. Thorat \cite{Thorat2020} had similarly critiqued Facebook's \textit{Free Basics} initiative, arguing that

\begin{quote}
    \textit{Free Basics evokes paradigms of national development, modernization, and progressivism that are rooted in technoutopian narratives. At this nexus of corporate and state interest in digital infrastructure lies a technoutopian belief that social, political, and economic problems in the Global South can be resolved by technological advancement, and nationwide issues of inequality and disparity will be ameliorated if poor citizens have access to the Internet} 
\end{quote}

This illustrates how technologies initially adapted for local needs are later taken over by big tech companies, revealing uneven power dynamics in design \cite{10.1145/3630106.3658934, 10.1145/3637316, 10.1145/1753326.1753522}. Additionally, these systems impose and claim to empower people who may no longer be able to understand or utilize them due to said standardization. It exemplifies digital colonialism, as it ignores decolonial perspectives and promotes a \textit{foreign is great} ideology, leading to feelings of unworthiness \cite{oji2020digital} and, consequently, a sense of \textit{resigned acceptance}.



Additionally, as discussed in section \ref{appper}, many participants discussed their dependence on the inherently personal nature of WhatsApp, as they rely on their trusted circles for practicing infrastructuring. Through coercive professionalization, Meta neglects the value of immaterial labor, such as building trust and empathy, which are essential in the Indian business context \cite{berger2020doing}. This aligns with feminist critiques of labor, which argue that immaterial labor—often feminized and undervalued—is frequently dismissed as "non-real" labor, despite its significant contribution to success \cite{oksala2016affective, terranova2012free}. Societal ideals and Meta’s exploitation of users lead to a feeling that their immaterial labor is undervalued, despite its role in generating financial gains. This undermines the legitimacy of users' practices and exposes the coloniality inherent in legitimization through coercive professionalization \cite{wyrtzen2017}. In this context, coercive professionalization is evident in how Meta imposes a standardized, profit-driven framework that disregards the local, relational, and affective forms of labor essential to the users' success. By devaluing these personal forms of labor, Meta's actions exemplify the coercive nature of professionalization, forcing users into a model that erases the subjective and context-- dependent labor that underpins their practices-- and this is another characteristic of coercive professionalization.

Consequently, users' marginalization within this system is evident in their acceptance of disruptions as normal, reflecting the ongoing impact of coercive professionalization similar to the effects of digital colonialism \cite{Sturmer2021}. Many participants resigned themselves to the platform’s changes, viewing them as inherent challenges of operating in a digital space not designed for their needs. A severe consequence was the emergence of feelings of \textit{inadequacy and doubt}—as P11 noted, despite supporting her family through her business, she questioned, \textit{"Am I really a businesswoman?"} Similarly, P4 and P7 described their business activities as secondary or hobby-like, managing significant responsibilities within a framework that does not support their scale or needs. Despite this, they did not express dissatisfaction, believing they were not engaged in \textit{real business}. While the perceived lack of agency may stem from insufficient technological expertise and diverse cultural factors, it also reflects a rationalization of diminished self-worth due to the repossession \cite{doi:10.1177/1461444816629474} of infrastructure through coercive professionalization. This process again echoes colonial dynamics, where the imposition of external standards undermines local practices and reduces the professional agency of those involved \cite{10.1145/3274340}. 

Finally, in India and many other Global South cultures, self-promotion is often approached with considerable restraint, and individuals are typically less inclined to overtly display their achievements \cite{merkin2018individualism}. This understated approach is valued as a means of achieving more with less and is held in high regard. However, Meta's coercive professionalization is such that by imposing a professional framework that is not contextually aligned with those who originally built the business and whose needs it purports to address, it reinforces ideals of digital colonialism that marginalize and devalue local practices \cite{10.1145/3274340}.

\subsection{Implications for Design}

The findings emphasize the need to include decolonial perspectives when analyzing how platforms are infrastructuralized through appropriation. This aligns with prior research that ask to \textit{make visible the power dynamics and influence of Western European standardization processes on knowledge-making and communication practices} \cite{10.1145/3328020.3353927, 10.1145/3283458.3283497}. Additionally, for cultures that are not normative, and undocumented within \textit{western rigor}, designing systems that show \textit{forms of user-centred design that strive to fit technologies to a stabilized notion of the user} should be questioned \cite{doi:10.1177/0162243910389594, 10.1145/2662155.2662195}.


Decolonizing infrastructure requires systems to adapt to technological changes in ways that respect and accommodate diverse cultural needs \cite{schaefer2021understanding}. While efficiency and automation are important, this study emphasizes that platforms should focus on stabilizing systems that truly support local communities, rather than transforming them into something unfamiliar, like WhatsApp Business. Developments should enhance, not obscure, the tools communities appropriate and rely on. This approach aligns with participatory design, involving stakeholders in the process rather than imposing generic, one-size-fits-all solutions \cite{herbst2023developing}. Drawing from the findings in this study, I outline methods to support and encourage successful appropriation. 


Firstly, WhatsApp could maintain and enhance core personal features that small business owners rely on. For instance, users like P8 effectively utilize personal communication tools, such as status updates and group chats, for business purposes. Therefore, WhatsApp (and other platforms) could consider implementing a \textit{"Legacy Mode"} that allows users to retain and access familiar features while integrating new ones, thus supporting a culture of appropriation and tailoring \cite{DiSalvo2013, 10.1145/97243.97271}. Users could have the option to keep using their personalized WhatsApp setup, with that setup being recognized as legitimate. This could be made possible if WhatsApp allowed users to choose and verify the version they want to keep, giving them official approval for their customized setup.

Additionally, simplifying user interface updates is crucial. As P4 and P7 noted, frequent updates can be confusing and time-consuming. WhatsApp could introduce a \textit{"What’s New"} section that provides concise summaries and visual guides in local languages. This feature could reduce the learning curve and associated frustration, while also fostering professional agency \cite{10.1145/3274340, 10.1145/97243.97271, wulf2008component}. By presenting updates in digestible formats with localized visual cues, WhatsApp could lower the cognitive burden on users, particularly those with limited technical expertise \cite{oviatt2006human}, thereby facilitating smoother transitions with each update.

Furthermore, fostering collaborative learning could significantly benefit small business users \cite{doi:10.1177/1461444816629474}. When it comes to communities based on communal interests, values, or goals, collaborative learning can be particularly powerful, as these communities often have strong shared ties that can foster deeper, more meaningful interactions \cite{panitz1998encouraging}. P3 and P8 highlighted how they relied on personal networks to manage changes. WhatsApp could formalize this by creating support forums or user communities where small business owners can share experiences and solutions \cite{10.1145/97243.97271, wulf2008component}. This would facilitate peer-to-peer learning and provide a platform for collective problem-solving. 

Lastly, WhatsApp could reconsider its approach to monetization, particularly for small businesses, as rising costs and complexity can be significant barriers, as seen with users like P3 and P13. Implementing tiered pricing for small businesses could help reduce financial burdens and support long-term growth—a practice common in developing countries within pharmaceutical industries \cite{moon2011winwin}. This approach could enhance adaptability by allowing users at different business levels to pay for a version that they need, as opposed to each feature, as scalability needs of different users may vary \cite{4085532}. Instead of charging per conversation or per message, a tiered system could offer plans based on the capabilities utilized, providing a more holistic pricing model that aligns with the specific needs of each business.

%% file: limitations.tex
\section{Limitations and Future Work}


This study has limitations, including a small sample size, which is common in qualitative HCI research, especially in critical cultural contexts \cite{eisenhart2009generalization}. Consequently, it does not aim to generalize findings but rather to elucidate the unique practices of platform users and their role in the infrastructural development of profit-oriented companies \cite{au2019thinking}. Moreover, there is a potential skew towards women in the demographics because, in many Global South cultures, women are more likely to use messaging apps extensively, while men often engage in traditional jobs. This pattern can affect the data, as it reflects broader cultural and economic roles rather than indicating a limitation of the study itself \cite{kantor2002sectoral}. Additionally, the interviews were conducted while I was situated in a different country from the participants \cite{Tuyisenge2023, Tarrant2013}. Despite sharing similar cultural backgrounds and experiences, this geographical separation may have affected the establishment of trust and ongoing communication, potentially influencing the participants' willingness to fully disclose information \cite{mullings1999insider}. However, the recruitment process involved multiple rounds of purposeful sampling, with many participants being recruited through personal connections and referrals. This method is noted to bring a sense of \textit{word-of-mouth} trust, which likely encouraged more open and honest discussions \cite{friedman2015qualitative}. 

Future work should focus on two key studies: First, a digital ethnographic study within WhatsApp groups of marginalized sellers to explore how they appropriate the platform for business communication, identifying infrastructural needs and challenges from within-- that could supplement results in this work to explicate actual features that may be useful for the community. Second, a feasibility analysis of WhatsApp and similar messaging applications to assess their accessibility, usability, and potential for adaptation, with the goal of refining these platforms to be user-friendly and effective for individuals with limited technical literacy. Each of these studies could also formulate strategies to support the appropriation of digital tools for innovative local uses, potentially exploring ways in which this appropriation can be legitimized.


%% file: conclusionandfuture.tex
\section{Conclusion}

I interviewed 14 WhatsApp users from India who used WhatsApp to manage their home-based small businesses. My research explored how they 1) appropriated WhatsApp for business purposes, 2) adapted to changes in WhatsApp Business, and 3) perceived these changes. I found that users frequently leveraged their personal contacts and connections for business, translating their personal use of WhatsApp into professional contexts. Additionally, they struggled to keep up with rapid changes and often felt that such platforms were not designed with their needs in mind. Many participants developed workarounds for changes that did not suit them. Lastly, I observed a diminished sense of self-esteem related to the ongoing processes of appropriation and infrastructuring. I contributed to HCI research in several ways. Firstly, I introduced the concept of \textit{coercive professionalization}, which describes how Meta has sought to co-opt infrastructural platforms initially developed through local community appropriation. This process represents a form of reverse appropriation, where these platforms are reclaimed by big tech companies. I also explored the features of coercive professionalization in how they relate to coloniality and marginalization. Finally, I discussed designs that can support the legitimization of local customization and appropriation of systems.